\documentclass[twocolumn,eqsecnum,pre]{revtex4}
\usepackage{bm}
\usepackage{graphicx}
\usepackage{amsmath}

\newcommand{\figuresatend}{}

\newcommand{\sinf}{\sigma_{\infty}}
\newcommand{\pinf}{\phi_{\infty}}
\newcommand{\avib}{a_{\text{vib}}}
\newcommand{\pvib}{P_{\text{vib}}}
\newcommand{\pis}{P_{\text{IS}}}
\newcommand{\phistar}{\phi^*}
\newcommand{\phimin}{\phi_{\text{min}}}
\newcommand{\phimax}{\phi_{\text{max}}}
\newcommand{\pderiv}[2]{\frac{\partial #1}{\partial #2}}
\newcommand{\pderivtext}[2]{\partial #1 / \partial #2}
\newcommand{\pderivfull}[3]{\left( \frac{\partial #1}{\partial #2} \right)_{#3}}

\newcommand{\deriv}[2]{\frac{d #1}{d #2}}
\newcommand{\derivtext}[2]{d #1 / d #2}

\newcommand{\ginf}{\gamma_{\infty}}
\newcommand{\gmin}{\gamma_\text{min}}
\newcommand{\bg}{\beta_\text{IG}}
\newcommand{\tg}{T_\text{IG}}
\newcommand{\rs}{\rho_\text{S}}
\newcommand{\tk}{T_\text{K}}
\newcommand{\ane}{a_\text{NE}}

\newcommand{\bint}{\beta_\text{int}}
\newcommand{\bquench}{\beta_\text{q}}

\begin{document}

\makeatletter
\floats@sw{\newcommand{\mycaption}[1]{\caption{#1}}}
{\setlength{\abovecaptionskip}{2 in}\newcommand{\mycaption}[1]{\caption[#1]{Shell, et. al.}}}
\renewcommand{\l@figure}[2]{\figurename\ #1\\[2cm]}
\makeatother

\title{Energy landscapes, ideal glasses, and their equation of state}

\author{M. Scott Shell}
\email[email:]{shell@princeton.edu}
\author{Pablo G. Debenedetti}
\email[corresponding author; email:]{pdebene@princeton.edu}
\affiliation{Department of Chemical Engineering\\Princeton University, Princeton, NJ 08544}
\author{Emilia La Nave}
\email[email:]{emilia.lanave@phys.uniroma1.it}
\author{Francesco Sciortino}
\email[email:]{francesco.sciortino@phys.uniroma1.it}
\affiliation{Dipartimento di Fisica and Istituto Nazionale per la Fisica della Materia\\Center  for Statistical Mechanics and Complexity and UdR,\\
Universit\'{a} di Roma La Sapienza, P.le Aldo Moro 2, I-00185 Rome, Italy}
\date{December 19, 2002}

\begin{abstract}
Using the inherent structure formalism originally proposed by Stillinger and Weber [Phys. Rev. A \textbf{25}, 978 (1982)], we generalize the thermodynamics of an energy landscape that has an ideal glass transition and derive the consequences for its equation of state.  In doing so, we identify a separation of configurational and vibrational contributions to the pressure that corresponds with simulation studies performed in the inherent structure formalism.  We develop an elementary model of landscapes appropriate to simple liquids which is based on the scaling properties of the soft-sphere potential complemented with a mean-field attraction.  The resulting equation of state provides an accurate representation of simulation data for the Lennard-Jones fluid, suggesting the usefulness of a landscape-based formulation of supercooled liquid thermodynamics.  Finally, we consider the implications of both the general theory and the model with respect to the so-called Sastry density and the ideal glass transition.  Our analysis shows that a quantitative connection can be made between properties of the landscape and a simulation-determined Sastry density, and it emphasizes the distinction between an ideal glass transition
and a Kauzmann equal-entropy condition.
\end{abstract}

\maketitle
\section{Introduction}
\label{introduction}

Deeply supercooled liquids and glasses occupy a prominent place in
modern science and engineering.  In addition to their established and
important presence among polymeric materials, glasses are
becoming key elements in new technologies, such as pharmaceutical
preservation and corrosion-resistant alloys \cite{932}.  The typical
route to the vitreous state is to supercool a liquid fast enough so
that crystallization is thwarted; hence the properties of 
glasses are intimately linked to those of the metastable liquids from
which they are made \cite{1459}.  At sufficiently low
temperatures during the cooling process, the structural relaxation
times of the metastable liquid become so slow that its mechanical
properties begin to resemble those of a solid and the material is
no longer in equilibrium relative to a laboratory time scale.  Though
at this point its mechanical behavior is solid-like, the
distinguishing structural feature of a glass is that it possesses no 
long-range microscopic order.

The major industrial prominence of glasses tends to belie our rather limited
theoretical understanding of these materials.  Although significant
progress has occurred in recent years \cite{goetze,mezpar,speedy,wolynes}, 
many questions still remain
concerning the appropriate thermodynamic treatment of the glassy state 
\cite{teo,kurchan,franz}, 
the relationship between kinetics and thermodynamics in these systems 
\cite{697,1413,953,547,914,1785,1645},
and the connection between molecular architecture and macroscopic behavior 
\cite{932,1790}.  The present work
addresses the first of these topics and aims to clarify some of the most
conspicuous questions about the thermodynamics of supercooled liquids
and glasses.  The unresolved issues in glass thermodynamics are quite varied, and
an important goal of this work is to investigate them with a common theoretical
framework capable of complete thermodynamic description of the
liquid state.  We propose a simple landscape-based equation of state for a 
supercooled liquid and demonstrate the usefulness of this approach by comparing
theoretical predictions with simulation data.

One unresolved issue regards the so-called Kauzmann temperature, $\tk$
\cite{1578, 932, 1459, 745}.  As a liquid is supercooled, its larger
heat capacity relative to the crystal causes the melting entropy to be
gradually consumed until it appears that at this distinguished temperature, the
liquid and solid entropies become equal.  Upon further cooling below
the Kauzmann temperature, the liquid would eventually attain a negative entropy
and would hence appear to violate the third law of thermodynamics.
Experimentally, however, the Kauzmann temperature must
be extrapolated because upon cooling the glass transition
intervenes at higher temperature, thus preventing any such violation.
Nevertheless for many liquids, the extrapolation needed to attain the
equal-entropy condition is quite modest, and attempts to understand the
nature and implications of this impending entropy crisis underlie the
thermodynamic interpretation of the glass transition.

The idea that a kinetically-controlled glass transition prevents a
thermodynamic catastrophe seems rather unsettling and has led to the
notion of a thermodynamic ``ideal'' glass transition at $\tk$
\cite{745}.  In this sense, the experimentally observed glass
transition is viewed as a kinetically-blurred manifestation of an underlying
phase transition.  The ideal glass transition has come to
be associated with the sudden entrapment of a system in a
lowest-energy, unique amorphous configuration accessed in the limit of
infinitely slow cooling.  It is now recognized that in this
definition, an ideal glass transition can actually be a rather
distinct occurrence from a Kauzmann equal-entropy point; the former is defined by
the amorphous state alone, whereas the latter makes reference to the
crystal \cite{1644,sktjpcm}.

The energy landscape formalism introduced by Stillinger and Weber has
been an important theoretical tool for formulating a thermodynamics of
glasses \cite{801}.  From this perspective, a liquid is described by
the structurally distinct configurations through which it evolves,
each termed an inherent structure, plus the kinetic ``vibrational''
distortions around these configurations.  By definition, the inherent
structures are the complete collection of mechanically stable particle
packings (local potential energy minima), and any one can be found
from a given configuration by energy minimization.  This permits a
rigorous theoretical separation of liquid-state properties into
inherent structure and vibrational components, including the energy,
entropy, and pressure.
The energy landscape paradigm has greatly facilitated the
understanding of low-temperature liquids and their glasses \cite{937,
932, 1643, 1641, 1645, 914, 936, 953, 1654, mossa1,mossa2}.

In the context of the energy landscape, ideal glasses acquire the
rigorous definition of zero \emph{configurational} entropy \cite{745}.
The configurational entropy is that part of the entropy associated with the degeneracy of
inherent structures of a given potential energy; presumably when the
temperature is low enough, the system samples only its minimum-energy
amorphous conformation and its configurational entropy vanishes.  In
contrast, Kauzmann points are defined by the equality of total crystal
and liquid entropies.  Since the configurational entropy of a crystal
is zero, the difference between an ideal glass transition and a
Kauzmann point is due solely to differences in vibrational entropies between
the supercooled liquid and the stable crystal.
In fact, it has been observed that several
real substances do in fact exhibit Kauzmann equal-entropy points
without violation of the third law, and in these systems the
contribution of vibrational entropies is essential to the existence of
a Kauzmann condition \cite{1644}.  On the other hand, theoretical
arguments have challenged the possibility of an ideal glass transition
at finite temperature by examining the effects of elementary
excitations on the configurational entropy \cite{745}.
Still, whether an ideal glass exists in supercooled liquids as a
thermodynamic phenomenon underlying the laboratory (kinetic)
glass transition remains an important open question.
  
Our objective is to provide a simple landscape-based thermodynamic framework
for liquids, including their equation of state, and apply it to the 
investigation of some of these questions.  We make no attempt to describe
nonequilibrium states in our theory, such as the kinetically arrested glasses
observed in experiment, but instead invoke the possibility of an
ideal glass transition.  In Section \ref{phenomenology} we briefly review some of the empirical 
observations made from computer simulations about the relationship between
the equation of state and the energy landscape.  We proceed in Section
\ref{Theory} to outline our theory and examine some of its
implications for these observations as well as for the notion of an
ideal glass transition.
We then describe in Section \ref{model} what is perhaps the
simplest quantitative model for an energy landscape of a simple
liquid.  Finally, in Section \ref{conclusions} we remark on the
results that this equilibrium theory has for supercooled liquids and note
areas of future research in light of them.

\section{The equation of state in the landscape paradigm}
\label{phenomenology}

Within the energy landscape formalism, it is possible to separate the equilibrium pressure of a liquid into contributions from its inherent structures and those due to vibrational displacements about the energy minima.  The component of the pressure due to inherent structures, $\pis$, is particularly convenient to study because it can be measured directly by computer simulation.  To calculate the inherent structure pressure at a given density and temperature, a large number of configurations are taken periodically from a molecular dynamics trajectory and their energy is minimized; this procedure locates the corresponding inherent structures, and the pressure of the minimized configurations is then calculated from the standard virial expression.  Based on a number of investigations which have used such a protocol, $\pis$ has emerged as an important feature of supercooled liquids \cite{547,1641,657,667,664,1782,520}.  In particular, it has suggested an unexpected connection between the liquid spinodal, microscopic heterogeneity, and the mechanical strength of materials \cite{657,547,1644,1782}.

\begin{figure}
\includegraphics[width=3.375 in]{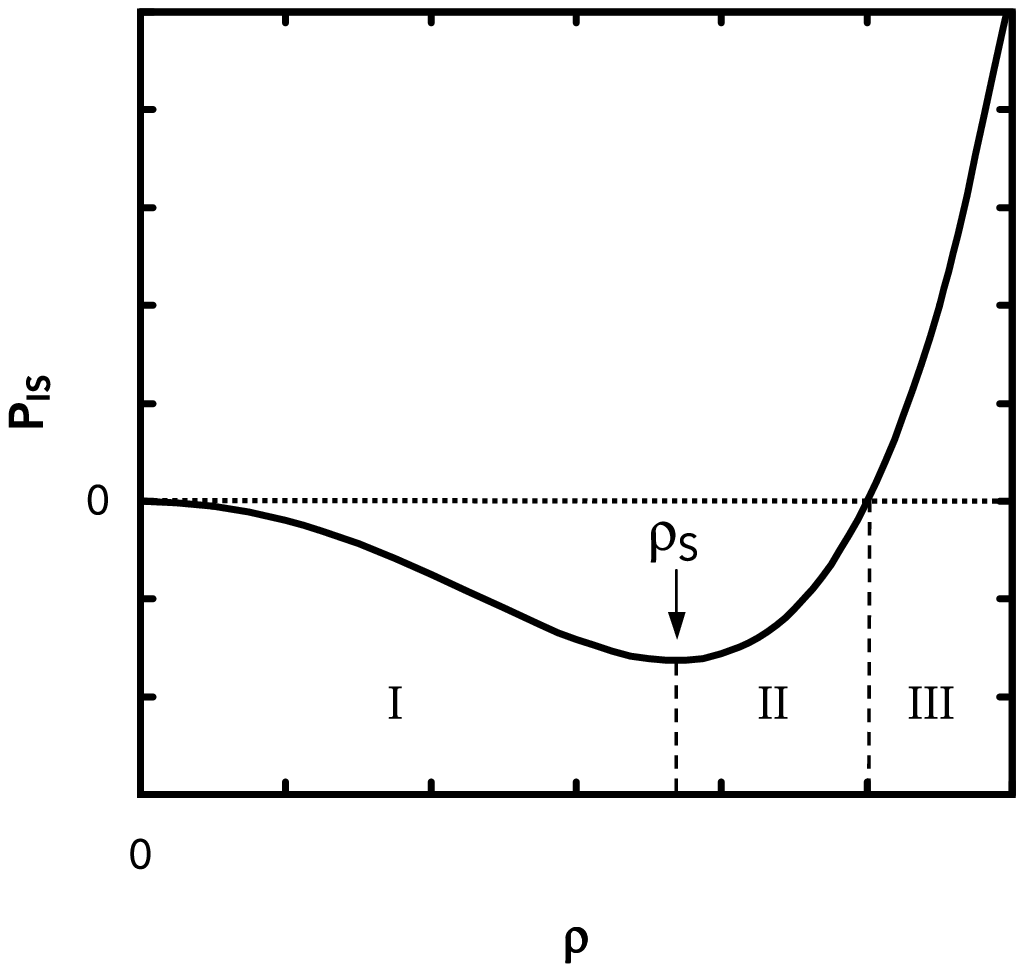}
\mycaption{Schematic of the density dependence of the inherent structure pressure, $\pis$, for simple fluids.}
\label{ljsastry}
\end{figure}

The inherent structure pressure has been studied in computer
simulations of several substances, and in all cases the density
dependence of $\pis$ has given rise to a picture qualitatively similar
to that shown in Fig. \ref{ljsastry} \cite{547, 667, 664, 657, 1782}.  It
has been found in simple liquids that, above the melting line, the
temperature of the equilibrium system from which the inherent
structures are generated has very little effect on the type of curve
shown in Fig. \ref{ljsastry},
meaning that $\pis$ is approximately independent of $T$ \cite{547}.
However, for all the systems studied so far below the melting temperature
(and for more complex systems such as water even above the melting 
line), there is a detectable influence of the temperature \cite{667,1641}.  

Fig. \ref{ljsastry} reveals the presence of qualitatively distinct
density regimes in $\pis$ \cite{547}.  As the density decreases from
region III to II, the system's inherent structures evolve from a state
of positive pressure to one of tension.  The density at which the
minimum in pressure occurs, the so-called Sastry density, $\rs$,
represents the point of maximum attainable tension.  At densities
below this point (region I), the system fractures and its inherent
structures are no longer homogeneous but instead contain significant
void regions \cite{547}.  The Sastry density is therefore the density
of limiting mechanical stability.  In the case of glassy substances,
whose mechanical properties are dominated by the slow relaxation from
one inherent structure to another, the Sastry density offers an
important measure of their ultimate strength
\cite{547,667,664,657,1782}.

The study of the thermal and mechanical stability boundaries of supercooled liquids has received 
considerable attention in recent years \cite{520,1644,sastry2,speedy2}.
In a number of investigations, it has been found that the Sastry
density, determined from finite-temperature simulations, is nearly
indistinguishable from the zero-temperature limit of the liquid branch
of the spinodal curve, determined from accurate equations of state \cite{520,1644}.  
These results imply a very weak dependence of the Sastry
density on the temperature of the equilibrated liquid from which it
is calculated.  Based on such results it has been proposed to
interpret $\rs$ as the zero-temperature terminus of the liquid
spinodal \cite{520}.  Under this hypothesis, the limit of
mechanical stability of inherent structures coincides with the limit
of thermodynamic stability of the liquid in the absence of thermal
motion.  This proposal suggests a particularly striking connection
between features of a liquid at high temperatures and absolute zero;
however, the underlying ideas are controversial \cite{sastry2} and they have yet to
benefit from a rigorous formulation \cite{520,1644}.  A particularly
prominent criticism of the spinodal connection stems from
simulation results for a model glass-former in which the limit of
mechanical stability for the liquid was explicitly calculated; 
in this counter-example the zero-temperature
limit of liquid mechanical stability did not appear to converge to the Sastry 
density \cite{sastry2}.

This picture has become more intriguing with the recent observation in several simple models that the liquid spinodal and Kauzmann curve converge in the zero-temperature limit \cite{520,1644}.  If the connection between the spinodal and Sastry density is made, this means that $\tk(\rs)=0$.  The qualitative explanation for this occurrence offered in \cite{1644} is that the inherent structures at $\rs$ reach a point of zero configurational entropy, and hence represent an ideal glass, due to the constraint that they must be devoid of weak spots that would initiate fracture.  That is to say, the requirements of homogeneity and maximum tension limit the system to a single inherent structure at $\rs$.  This explanation invokes the notion of a system with zero configurational entropy; for this to apply to a Kauzmann point as well, the vibrational entropies of the crystal and metastable liquid must be equal.  Though this is the likely behavior at absolute zero, more insightful theoretical or quantitative arguments substantiating the Kauzmann-spinodal connection have not been offered.  Furthermore, detailed calculations for one particular model glassy system seem to exclude the zero-temperature spinodal-Kauzmann convergence \cite{sastry2}, questioning the universality of this observation.

\section{Theoretical framework}
\label{Theory}

\subsection{Free energy}

In the energy landscape formalism, the starting point in the analysis of a liquid is its potential energy surface (PES), the many-particle potential energy function of the configurational coordinates \cite{801}.  For a single-component system of $N$ structureless particles, this surface exists in a space of $3N+1$ dimensions.  In this enormous dimensionality, the PES contains an overwhelmingly large number of local minima, the inherent structures, each corresponding to a mechanically stable packing of the particles \cite{673}.  By following a steepest descent trajectory, every configuration can be uniquely mapped onto one of these minima.  Consequently, the entire configuration space can be tiled into basins around inherent structures.  The collection of configurations that corresponds to a given inherent structure is called its basin of attraction.  This approach allows a rigorous separation of the partition function into contributions from inter-basin movement and vibration within a basin \cite{801}:
\begin{equation}
Z=e^{-\beta A} = C \int_{\phimin(\rho)}^{\phimax(\rho)}{e^{ N \left[ \sigma(\phi,\rho) - \beta \phi - \beta \avib (\beta, \phi, \rho) \right]} d\phi}
\label{basinZ}
\end{equation}
where $Z$ is the canonical partition function, $\beta$ ($=1/{k T}$) is the Boltzmann factor, $A$ is the Helmholtz free energy, $C$ is a constant with units of reciprocal energy, $\phi$ is the potential energy per particle, $\rho$ is the number density, $\sigma(\phi,\rho)$ is the basin enumeration function, and $\avib (\beta, \phi, \rho)$ is the basin vibrational free energy.  In this expression, $\avib$ gives the per-particle free energy when the system is confined to an average basin of depth $\phi$, and the basin enumeration function is defined such that $C \exp{\left[ N \sigma(\phi, \rho) \right]d\phi}$ gives the number of inherent structures with potential energy per particle $\phi \pm \frac{1}{2}d\phi$.  For the supercooled liquid, both the vibrational free energy and basin enumeration function account only for those inherent structures and basins which are sufficiently devoid of crystallites.  The partition function $Z$ therefore corresponds to a restricted ensemble for the metastable liquid \cite{745}.

The integral in Eq.\ \ref{basinZ} is performed over the per-particle potential energy of inherent structures and ranges from the lowest-lying amorphous configuration (with energy $\phimin$) to the least stable arrangement (with energy $\phimax$.)  Presumably at these extreme energies, there exists a single, unique amorphous basin; it therefore follows that $\sigma(\phimin)=\sigma(\phimax)=0$.  The condition of zero configurational entropy does not exclude the existence of inherent structures with sub-exponential system size dependence.  Consequently, the notion of a unique amorphous inherent structure at $\sigma=0$ refers to the subset of basins that are thermodynamically relevant in the large-system limit.  At this point, we should note that in the following discussion, the notation $\partial{\avib(\beta,\phistar,\rho)}/\partial{\phi}$ is used to indicate the $\phi$-derivative of $\avib$ taken at constant values of its remaining natural variables, $\beta$ and $\rho$, and \emph{evaluated} at $\phistar$.

The transformation in Eq.\ \ref{basinZ} offers enormous simplification in the analysis of liquids; given $\sigma$ and $\avib$, for which good theoretical functionalities can be written, the usual $3N$-dimensional integral over positions in the configurational partition function becomes a one-dimensional integral over potential energy.  Furthermore, Eq.\ \ref{basinZ} permits the separate determination of vibrational and basin contributions.  Often the vibrational component can be successfully modeled using a harmonic approximation, while basin degeneracies can be understood through packing considerations \cite{914, 936, 953, 673, 659}.  This casting of the partition function offers a particularly convenient framework for deeply supercooled liquids, in which intrabasin equilibration occurs on much shorter timescales than interbasin hopping \cite{935}.  

We focus our attention on the implications of the PES for supercooled liquids and ideal glasses.  In the large system limit, the integral in \ref{basinZ} will be dominated by a maximum value of the exponential term at $\phi=\phistar$.  This means that, at the given temperature and density, the system will sample configurations whose overwhelming majority have energy $\phistar$; thus one can consider only those contributions corresponding to $\sigma(\phistar,\rho)$ and $\avib(\beta, \phistar, \rho)$:
\begin{eqnarray}
Z & \approx & e^{ N \left[ \sigma(\phistar,\rho) - \beta \phistar - \beta \avib (\beta, \phistar, \rho) \right]}
\nonumber
\\
A/N & \approx & \phistar - kT\sigma(\phistar,\rho) + \avib(\beta,\phistar,\rho)
\label{basinZinfinite}
\end{eqnarray}
where there is a clear distinction between the basin ($\phistar-kT\sigma$) and vibrational ($\avib$) contributions to the free energy.  In particular, $Nk\sigma$ gives the configurational entropy, that is, the entropy due to the system's exploration of multiple basins of a given depth.  This is consistent with conventional notions of entropy; when the system is confined to a single basin, as in a crystal, the configurational entropy vanishes \cite{680}.

The condition that determines the mean inherent-structure energy $\phistar$ requires special attention.  In the usual approach, the maximum of the integrand in \ref{basinZ} is located via the derivative of the exponential term \cite{1643},
\begin{eqnarray}
\pderiv{\sigma(\phistar,\rho)}{\phi} & = & \beta +\beta \pderiv{\avib(\beta,\phistar,\rho)}{\phi}
\nonumber \\
& = & \frac{\beta}{B(\beta,\phistar,\rho)}.
\label{derivmax1}
\end{eqnarray}
This is an implicit expression for $\phistar$ as a function of $\beta$ and $\rho$.  Here we have introduced the function $B$ for notational convenience; it appears frequently in this theory and is given by
\begin{equation}
B(\beta,\phi,\rho) \equiv \left[ 1 + \pderiv{\avib(\beta,\phi,\rho)}{\phi} \right]^{-1} .
\label{Bdef}
\end{equation}
For simplicity in this analysis, one might assume the vibrational free
energy to be independent of inherent structure energy, i.e.,
$B(\beta,\phi,\rho) = 1$.  This is the rigorous limit of $B$ at absolute
zero, where the vibrational free energy and its $\phi$-derivative vanish.
At finite temperatures, however, this approximation gives meaningful results
only when the curvature of basins is weakly dependent on their depth,
a condition that might be expected for very deep basins (and low
temperatures) where the majority of the particles are ``well-packed.''
Still, we emphasize that this remains an approximation
as a nonzero dependence of basin shape on depth has been found in a 
number of systems \cite{1073,936,1175,francis}.

The condition described by Eq.\ \ref{derivmax1} holds for all
temperatures if the slope of the basin enumeration function tends
toward infinity at $\phimin$.  With such behavior, the mean inherent
structure energy decreases continuously to $\phimin$ as the
temperature approaches zero (i.e., as $\beta \to \infty$.)  For basin
enumeration functions with a finite slope at their minimum energy, the
inherent structure energy reaches $\phimin$ at a nonzero temperature.
At lower temperatures, the value $\phistar$ which produces the
integrand maximum will always occur at the low-energy extremum.  This
condition, in which the system is trapped in a unique amorphous 
minimum-energy configuration (or in one of a sub-extensive number thereof),
is the ideal glass (IG) transition.  The $\phistar$ condition is
then given by
\begin{equation}
\phistar \text{ s.t. } \left\{ \begin{array}{ll}
\pderiv{\sigma(\phistar,\rho)}{\phi} =
\frac{\beta}{B(\beta,\phistar,\rho)} & \text{for } \beta < \bg(\rho)
\\ \phistar = \phimin(\rho) & \text{for } \beta \ge \bg(\rho)
\end{array} \right.
\label{extrmax1}
\end{equation}
where $\bg(\rho)$ [$=1/k\tg(\rho)$] is formally determined by
\begin{equation}
\pderiv{\sigma(\phimin,\rho)}{\phi} = \frac{\bg}{B(\bg,\phimin,\rho)}.
\label{betaKdef1}
\end{equation}
In this case, a system in equilibrium at temperatures below $\tg$
would be confined to the lowest-lying basin available at that density,
with energy $\phimin$.  Accordingly, $\sigma=0$ for the entire range
of temperatures $T<\tg$.  In Eq.\ \ref{extrmax1}, it is important
to observe that with reasonable molecular interactions, the $\phi$
derivative of the vibrational free energy should remain bounded as $T
\to 0$.  Eq.\ \ref{extrmax1} is the general case for basin
enumeration functions.  In the limit $\pderivtext{\sigma(\phimin)}{\phi}
\to \infty$, we have that $\tg \to 0$ and the expression in
\ref{extrmax1} reduces to \ref{derivmax1}.

The preceding equations provide a thermodynamic framework for an ideal
glass ($\sigma=0$), motivated by the energy landscape formalism.  The
transition to this state at $\tg$ is marked by the sudden confinement
of the system to a unique, lowest-energy amorphous configuration, that
is to say, the configurational entropy becomes zero.  By our notation,
we strictly identify $\tg$ as an ideal glass transition, though
previous studies have named the point at which the configurational
entropy of the liquid becomes equal to the configurational 
entropy of the crystal a Kauzmann  
point \cite{953,sktjpcm,680,936,1643,932}, extending the
original definition of  ``Kauzmann transition.''  The
reader is reminded, however, that coincidence with a true Kauzmann
transition (i.e., as originally defined) occurs only when the vibrational entropies of the liquid
and crystal are equal.  To prevent any confusion, we avoid calling
the ideal glass transition line a Kauzmann line.
Further in this discussion we will explore the
extent to which the observations regarding Kauzmann points presented
in Sections \ref{introduction} and \ref{phenomenology} also hold for
ideal glasses, which will effectively be a test of the crystal-liquid
vibrational entropy equality.

\subsection{Equation of state}

It is apparent from the energy landscape formalism that there are two classes of basin enumeration functions \footnote{We do not consider
basin enumeration functions whose derivative is non-monotonic.  Such behavior would introduce first order phase transitions.}.  Those which have a finite slope at $\phimin$ give rise to an ideal glass at finite temperature; those whose slope is infinite lack such a transition.  To consider the implications of these two possibilities on the equation of state, we examine the volume derivative of Eq.\ \ref{basinZinfinite}.  In the ideal glass-free case, this is
\begin{eqnarray}
P & = & \frac{\rho^2}{\beta} \pderiv{\phistar(\beta,\rho)}{\rho} 
\left[ \beta + \beta \pderiv{\avib(\beta,\phistar,\rho)}{\phi} - \pderiv{\sigma(\phistar,\rho)}{\phi} \right]
\nonumber \\ 
& & + \rho^2 \pderiv{\avib(\beta,\phistar,\rho)}{\rho} - 
\frac{\rho^2}{\beta} \pderiv{\sigma(\phistar,\rho)}{\rho}
\nonumber \\
& = & \pvib(\beta,\rho) + \pis(\beta,\rho)
\label{noglassP}
\end{eqnarray}
where the term in brackets vanishes according to the equilibrium condition in Eq.\ \ref{derivmax1}, and $\pvib$ and $\pis$, the vibrational and inherent structure pressure contributions, are conveniently defined by adding and subtracting $(\rho^2 B / \beta) \pderivtext{\sigma}{\rho}$ to the last two terms in this equation and rearranging to obtain
\begin{eqnarray}
\pvib(\beta,\rho) & \equiv & \rho^2 \pderiv{\avib(\beta,\phistar,\rho)}{\rho}
\nonumber \\
& & + \frac{\rho^2}{\beta} \left[ B(\beta,\phistar,\rho) - 1 \right] \pderiv{\sigma(\phistar,\rho)}{\rho}
\label{pvibdef} \\
\pis(\beta,\rho) & \equiv & -\frac{\rho^2}{\beta} B(\beta,\phistar,\rho) \pderiv{\sigma(\phistar,\rho)}{\rho} .
\label{pisdef}
\end{eqnarray}
This separation of the pressure components is constructed so that the inherent structure pressure coincides with that measured in simulation studies of energy landscapes.  Perhaps more revealing, $\pis$ is equivalently defined by the negative volume derivative of average inherent structure energy at constant configurational entropy.  The equivalence with Eq.\ \ref{pisdef} is established by the mathematical identity
\begin{equation}
\pderivfull{\sigma}{\rho}{\phi} = -\pderivfull{\sigma}{\phi}{\rho}  \pderivfull{\phi}{\rho}{\sigma}
\label{cyclic}
\end{equation}
when used in combination with the equilibrium condition in Eq.\ \ref{derivmax1} and substituted in the pressure expressions in Eqs.\ \ref{pvibdef} and \ref{pisdef}:
\begin{eqnarray}
\pvib(\beta,\rho) & = & \rho^2  \bigg[ 
\pderiv{\avib(\beta,\phistar,\rho)}{\rho} 
\nonumber \\
& & + \pderiv{\avib(\beta,\phistar,\rho)}{\phi} 
\pderivfull{\phistar}{\rho}{\sigma} \bigg]
\label{pvibdef2}
\\
\pis(\beta,\rho) & = &  \rho^2 \pderivfull{\phistar}{\rho}{\sigma}
\label{pisdef2}
\end{eqnarray}
where the $\phistar$-derivative is taken along a path whose configurational entropy is consistent with the given density and temperature.  For a more detailed discussion of the separation of vibrational and configurational contributions to the pressure, the reader is referred to the appendix.

In the case of a finite-temperature ideal glass, Eq.\ \ref{noglassP} still describes the pressure for $T>\tg(\rho)$.  Below the ideal glass transition, however, both $\phistar$ and $\sigma$ are constant at fixed density, constrained to their value at $\tg$.  Using these constraints in the expression for the free energy and subsequently taking its volume derivative then gives the pressure below the ideal glass transition:
\begin{eqnarray}
P & = & \rho^2  \left[ \pderiv{\avib(\beta,\phimin,\rho)}{\rho} 
+ \pderiv{\avib(\beta,\phimin,\rho)}{\phi} \deriv{\phimin(\rho)}{\rho} \right]
\nonumber \\
& & + \rho^2 \deriv{\phimin(\rho)}{\rho}
\nonumber \\
& \equiv & \pvib'(\beta,\rho) + \pis'(\rho)
\label{glassP}
\end{eqnarray}
where $\pvib'$ and $\pis'$ correspond to the first and last terms, respectively, and the prime symbols indicate the formal difference in these definitions from those in Eqs.\ \ref{pvibdef} and \ref{pisdef}.  Despite dissimilar expressions above and below $\tg$, the individual pressure components remain continuous.  Along the ideal glass transition locus, $\tg(\rho)$, the configurational entropy remains constant at zero and $\phistar=\phimin$.  Considering this behavior in the alternate pressure expressions in Eqs.\ \ref{pvibdef2} and \ref{pisdef2}, we have the equalities
\begin{eqnarray}
\pvib(\bg,\rho) & = & \pvib'(\bg,\rho)
\label{pvibcontinuity} \\
\pis(\bg,\rho) &  = & \pis'(\rho) .
\label{piscontinuity}
\end{eqnarray}
It is important to note here that the inherent structure pressure has no temperature dependence below the ideal glass transition, as is also the case with the inherent structure energy.  Using the equivalence of pressure expressions at $\tg$, one can write the inherent structure pressure for $T \le \tg$ in terms of its properties at the ideal glass transition:
\begin{equation}
\pis'(\rho) = -\frac{\rho^2}{\bg} B(\bg,\phimin,\rho) \pderiv{\sigma(\phimin,\rho)}{\rho}.
\label{pisglassP2}
\end{equation}

With the equation of state in \ref{glassP}, we can immediately evaluate the Sastry density, which we now define in a precise way as the zero-temperature minimum in the inherent-structure pressure as a function of density,
\begin{equation}
\rs \deriv{^2\phimin(\rs)}{\rho^2} + 2 \deriv{\phimin(\rs)}{\rho} = 0 .
\label{sastryeqn}
\end{equation}
The implication of this expression is that Sastry density is completely
determined by the density dependence of the energy of the lowest-lying basin.
This is not all that surprising, since one would expect $A/N \to
\phimin$ as $T \to 0$ and $\sigma(\phimin,\rho)=0$.  Eqs.\ \ref{pisdef} and \ref{glassP} also
provide the $T>0$ behavior of the pressure minimum and a corresponding
generalization to a finite-temperature Sastry density, which allows
the correspondence with computer simulation studies.  
We note that the Sastry density defined by Eq.\ \ref{sastryeqn} 
also coincides with the density at which the inherent
structure pressure along the ideal glass transition locus has a minimum, and
is consistent with the zero-temperature minimum of the total pressure,
where any contribution from the vibrational free energy vanishes.  

\subsection{Basin enumeration functions}

To understand more specifically the implications of a liquid's PES, we now turn to a class of simple expressions for the basin enumeration function which have generated considerable interest \cite{1643, 1641, 1175, 936}.  Such generic functions must of course be taken as approximate if they are to be universally applicable.  Nonetheless, these expressions provide a starting point for the analysis of supercooled liquids and have even produced quantitatively accurate models in several simulation studies \cite{936,1645}.  We consider basin enumeration functions of the general form
\begin{eqnarray}
\sigma(\phi,\rho) = \sinf(\rho) f[u] & \text{ with } & u \equiv \frac{ \phi - \pinf(\rho) }{\alpha(\rho)}
\label{genericdef}
\end{eqnarray}
where $f[u]$ is a dimensionless function varying between 0 and 1, $\sinf$ is the maximum value of the basin enumeration function, and $u$ is the energy parameter made dimensionless by the inherent structure characteristic energy range $\alpha$ and offset by an energy $\pinf$.  We define the parameters $\pinf$ and $\alpha$ such that $u=-1$ for $\phi=\phimin$ and $\pinf$ is the energy at the maximum value of the basin enumeration function (i.e., $\alpha = \pinf - \phimin$).  For distributions whose energy range is symmetric about $\pinf$, we then have $-1 \le u \le 1$ ($\phimin \le \phi \le 2\pinf - \phimin$).  Furthermore, the form in \ref{genericdef} gives the total number of basins from $\exp{(N\sinf)}$ \cite{745}.  

The above approach allows the ``extraction'' of generic properties of the basin enumeration function, such as height, width, and mean, from its specific functional form, given by $f[u]$.  One of the simplest and most commonly used models is the Gaussian landscape \cite{1643}, given by 
\begin{equation}
f_\text{Gaussian}[u]=1-u^2 .
\label{gaussian}
\end{equation}
As evident from this expression, the Gaussian form of the basin enumeration function has a finite slope at its energy minimum, and therefore, always leads to an ideal glass.

For basin enumerations of the form in \ref{genericdef}, equilibrium is determined by
\begin{equation}
u^*(\beta,\rho) \text{ s.t. } 
\left\{ \begin{array}{ll}
\deriv{f[u^*]}{u} = \frac{\alpha(\rho)\beta}{B(\beta,\phistar,\rho) \sinf(\rho)} & \text{for } \beta < \bg(\rho) \\
u^* = -1 & \text{for } \beta \ge \bg(\rho)
\end{array}
\right.
\label{gendefequilib}
\end{equation}
where $u^*$ is the equilibrium value of $u$ and the glass transition temperature is given by
\begin{equation}
\frac{\alpha(\rho)\bg}{B(\bg,\phimin,\rho) \sinf(\rho)} = \deriv{f[-1]}{u} .
\label{gendefglass}
\end{equation}
When the vibrational free energy is independent of basin depth ($B=1$), a scaling relationship evolves from Eqs.\ \ref{gendefequilib} and \ref{gendefglass} between the temperature, total number of basins, and basin distribution breadth.  That is, the density dependence of the mean inherent structure energy is contained in a dimensionless temperature variable $\alpha\beta/\sinf$.  Calculating the inherent structure pressure contribution from Eqs.\ \ref{pisdef} and \ref{gendefequilib},
\begin{eqnarray}
\pis(\beta,\rho) & = & \rho^2 \left[\deriv{\pinf(\rho)}{\rho} + u^*\deriv{\alpha(\rho)}{\rho} \right]
\nonumber \\
& & - \frac{\rho^2 B(\beta,\phistar,\rho)}{\beta} \left[ f[u^*] \deriv{\sinf(\rho)}{\rho} \right]
\label{gendefP}
\end{eqnarray}
which holds above the glass transition.  As $T \to \tg$, the last group of terms in this equation vanishes since $f[u^*] \to 0$, and the first group on the right hand side becomes the same as that in Eq.\ \ref{glassP} by the definition of $\pinf$ and $\alpha$.  

The form in Eq.\ \ref{genericdef} has interesting implications
for the apparent coincidence of the Sastry density and the
zero-temperature Kauzmann point when the crystal and liquid
vibrational entropies are equal.  We assume first that we have
a functional form for the basin enumeration function which gives rise to
an ideal glass transition, i.e., $\derivtext{f[-1]}{u}$ is finite.
Then using the fact that in the $T=0$ limit the vibrational free energy is 
independent of basin depth and following
Eq.\ \ref{gendefglass}, the $\tg$-$\rs$ convergence implies
\begin{equation}
\frac{\alpha(\rs)}{\sinf(\rs)} 
\left(\deriv{f[-1]}{u}\right)^{-1} \to 0
\label{gendefsasglass}
\end{equation}
where $\rs$ is found from Eq.\ \ref{sastryeqn}.  This result
implies that either $\sinf$ diverges or $\alpha$ vanishes at the
Sastry density.  The first case appears unlikely; elementary arguments
about the shredding behavior of low-density inherent structures imply
that $\sinf$ should vary continuously across $\rs$ \cite{659}.  It
thus appears that the range of inherent structure energies, given by
$\alpha$, must shrink to zero at the Sastry density.  Qualitatively,
this seems plausible.  At smaller volumes, the system boundaries can
serve to stabilize poorly packed, high-energy inherent structures, but
as the density is lowered into a state of tension, this is no longer
the case and the energy of the least stable structure necessarily
decreases.  With the increasingly constraining requirement of
homogeneity, the maximum attainable inherent structure energy may be
forced to converge on the minimum as the Sastry density is approached,
that is, $\phimax \to \phimin$ at $\rs$.

\section{Elementary landscape model}
\label{model}

We now derive properties of a basic energy landscape for simple liquids which serves as a minimum description for their analysis in the landscape formalism.  We consider a simple model which is particulary convenient to analyze in the energy landscape paradigm.  This idealized system consists of structureless, spherically symmetric particles interacting in pairs through soft-sphere repulsive forces of the type $r^{-n}$ and each experiencing a density-dependent mean-field attraction \cite{520, 1644}.  The potential energy of a configuration of particles at a given density is therefore given by
\begin{equation}
U(r^{3N},\rho)=\sum_{i<j}{\epsilon\left(\sigma/r_{ij}\right)^n} - N a \rho
\label{ssmf}
\end{equation}
where $r^{3N}$ represents the positions of the particles, $\epsilon$ and $\sigma$ are the characteristic energy and length scales of the soft-sphere part, $n$ is the soft-sphere exponent which controls the degree of the repulsion, $r_{ij}$ is the distance between particles $i$ and $j$, and $a$, a positive number, is the mean-field parameter with units of $[\text{energy}]\times[\text{volume}]$.  There are in fact only two independent parameters in this expression, the soft-sphere and mean-field coefficients, $\epsilon \sigma^n$ and $a$.  This particular choice of the potential energy function is motivated by the observation that liquid structure is primarily determined by repulsive forces while attractive interactions can be successfully incorporated by a background potential field serving to hold the particles together \cite{1648, 1649}.  The advantage of this system is that the repulsive part of the potential energy in any configuration is determined by the dimensionless particle positions, $s=V^{-1/3}r$, in such a way that the volume dependence can be readily extracted:
\begin{eqnarray}
U/N & = & \epsilon \sigma^n V^{-n/3} N^{-1} \sum_{i<j}{s_{ij}^{-n}} - a \rho
\nonumber \\
& \equiv & \gamma(s^{3N}) \rho^{n/3} - a \rho
\label{dlessenergy}
\end{eqnarray}
where $\gamma(s^{3N})$, with units of $[\text{energy}]\times[\text{volume}]^{n/3}$, has been defined to give the portion of the soft-sphere term independent of volume, and is given by
\begin{equation}
\gamma(s^{3N}) \equiv \epsilon \sigma^n \frac{1}{N^{1+n/3}} \sum_{i<j}{s_{ij}^{-n}} .
\label{gammadef}
\end{equation}
Due to this kind of volume dependence, the nonideal soft-sphere part of the configurational partition function depends on a single scaling variable $z=(\beta\epsilon)^{3/n}\rho\sigma^3$ \cite{1647}.  This reduction in parameter space has made the soft-sphere/mean-field (SSMF) model attractive for studies of the liquid state \cite{520, 1644}.  Previous investigations using an equation of state based on reasonably accurate simulation data have shown in the model the presence of a Kauzmann locus and the intersection of this curve with the $T=0$ liquid spinodal \cite{520}.  We aim to extend the work in these studies by using the scaling embodied in Eq.\ \ref{dlessenergy} to develop a model energy landscape.

It is well known that the total number of inherent structures in the family of soft-sphere models is independent of density \cite{673, 659}.  This follows directly from the scaling effect of the volume on the potential energy and forces, which means that inherent structures are defined by steepest descent trajectories in the dimensionless particle positions, $s^{3N}$.  Extending this observation, the energies of the most and least stable inherent structure configurations must scale with density as $\rho^{n/3}$.  Therefore, the number of inherent structures within any fractional portion of this energy range must be density-independent.  The result is that the functional form of the basin enumeration function for soft-spheres is rigorously density-independent, that is to say, Eq.\ \ref{genericdef} is exact in this case.  (We should note here that these observations are also the limiting high-density case for more realistic pair potential functions which incorporate an inverse-power repulsion, such as the Lennard-Jones potential.  In such systems at high densities, the relevant configurations of the partition function sample only the repulsive part of these potentials.)  Finally, the appended mean-field term does not change the inherent structures or their number, but simply serves to shift their energy.  With these considerations, we can elaborate on some of the properties of the basin enumeration function:
\begin{eqnarray}
\phimin(\rho) & = & \gmin \rho^{n/3} - a \rho
\label{ssmfscale1} \\
\pinf(\rho) & = & \ginf \rho^{n/3} - a \rho
\\
\alpha(\rho) & = & (\ginf - \gmin) \rho^{n/3}
\\
\sinf(\rho) & = & \sinf
\label{ssmfscale4}
\end{eqnarray}
where $\gmin$ and $\ginf$ correspond to the values of $\gamma$ for configurations at the low energy extreme and maximum of the soft sphere basin enumeration function, respectively.

\begin{figure}
\includegraphics[width=3.375 in]{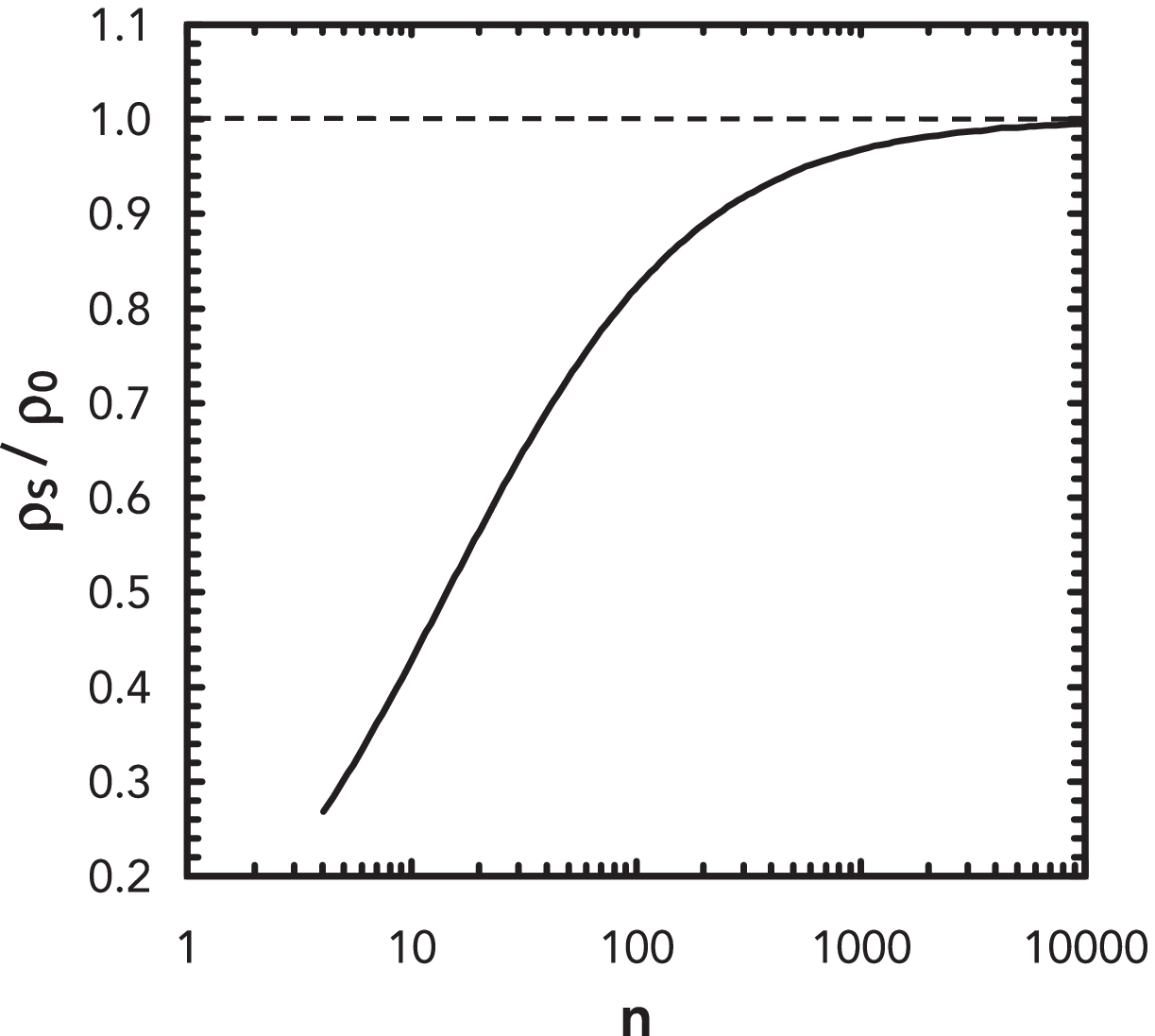}
\mycaption{Sastry density for the soft-sphere plus mean-field model as a function of repulsive exponent, $n$.  The limiting case for $n \to \infty$, shown by the dashed line, is $\rs/\rho_0 \to 1$.}
\label{ssmfsastryfig}
\end{figure}

Using expressions \ref{ssmfscale1}-\ref{ssmfscale4}, the Sastry density for the SSMF system follows directly from Eq.\ \ref{sastryeqn}:
\begin{equation}
\frac{\rs}{\rho_0} = \left[ \frac{1}{2} \left( \frac{n}{3} \right) \left( \frac{n}{3}+1 \right) \right]^{ \frac{1}{1-n/3} } .
\label{ssmfsastry}
\end{equation}
where we have introduced a characteristic density $\rho_0$, dependent on the soft-sphere exponent $n$ and defined by
\begin{equation}
\rho_0 \equiv \left( \frac{\gmin}{a} \right)^{ \frac{1}{1-n/3} }.
\label{rho0def}
\end{equation}
The significance of $\rho_0$ stems from Eq.\ \ref{ssmfsastry}; in the limit of hard sphere repulsive interactions for which $n \to \infty$, the Sastry density tends towards this reference density.  This trend is depicted in Fig. \ref{ssmfsastryfig}, which shows the evolution of $\rs$ over several orders of magnitude in $n$, starting with $n=4$.  Despite the existence of a zero-temperature pressure minimum, one important distinction between this model and ones incorporating more realistic attractive interactions is the absence of a distinguishing physical change at the Sastry density.  The SSMF system remains completely homogeneous below $\rs$ and does not exhibit the ``shredding'' behavior normally associated with this regime.  This makes it difficult in this case to justify, at least in a qualitative sense, the notion that the Sastry density represents a point of zero configurational entropy due to the constraint of homogeneity \cite{1644}.  Unfortunately, this type of justification requires a model of greater complexity, for which there is currently no obvious candidate.

In order to evaluate the equilibrium pressure in the SSMF fluid, we model the vibrational free energy in the harmonic approximation.  For consistency, we use the classical form \cite{1645, 914, 936, 953}:
\begin{eqnarray}
\avib(\beta,\phi,\rho) & \approx & \frac{3}{\beta} \ln{\left(\beta k\Theta_E\right)}
\label{avibapprox}
\\
\Theta_E & \equiv & \frac{h}{2\pi k} \sqrt{\frac{1}{m}\left<\deriv{^2 U}{r^2}\right>}
\label{thetae}
\end{eqnarray}
where $\Theta_E$ is taken as an Einstein temperature, $h$ is Planck's constant, $m$ is the particle mass, and the term in brackets is the geometric average of the eigenvalues of the Hessian matrix (i.e., the matrix of second derivatives of the potential energy with respect to particle coordinates.)  In this expression, there is an implicit dependence of $\Theta_E$ on basin depth and density, both whose effects manifest in the second derivative of the potential.  This means that for given values of $\phi$ and $\rho$, the Einstein temperature is essentially derived from a ``representative'' inherent structure of that basin depth at the specified density.  Making a transformation similar to that in Eq.\ \ref{dlessenergy}, we have
\begin{eqnarray}
\Theta_E & = & \frac{h}{2\pi k} \sqrt{\frac{\rho^{(n+2)/3}}{m} \left<N^{1/3}\deriv{^2 \gamma}{s^2}\right>}
\nonumber \\
& = & \frac{h}{2\pi k} \sqrt{\frac{\rho^{(n+2)/3}}{m} g(u)}
\label{thetae2}
\end{eqnarray}
where $g$, a function of the scaled basin depth $u$, gives the density-independent part of the second derivative term.  Essentially $g$ contains any dependence of basin curvature, and hence vibrational free energy, on basin depth; a $\phi$-independent $\avib$ simply implies that $g$ is constant.  Using the harmonic expression for the vibrational free energy in Eq.\ \ref{gendefequilib}, the ideal glass locus is given by
\begin{eqnarray}
\bg(\rho) & = & \frac{1}{\ginf-\gmin} \left[
\sinf \deriv{f[-1]}{u} - \frac{3}{2} \deriv{\ln{g[-1]}}{u} \right]
\rho^{-n/3}
\nonumber \\
& \equiv & C \rho^{-n/3}
\label{ssmfglass}
\end{eqnarray}
where we have condensed the constants in this expression into a single constant $C$ for clarity.  The equivalent expression for temperature is
\begin{equation}
\tg(\rho)= \frac{1}{C k} \rho^{n/3} .
\label{ssmfglassT}
\end{equation}  
The surprising consequence of this result is that an absolute zero ideal glass transition is reached only in the limit of infinitesimal density.  This means that the Sastry density, which is finite via Eq.\ \ref{ssmfsastry}, does not intersect the ideal glass curve at $T=0$.  This result may seem at odds with the results of Ref. \cite{1644} in which it was found from numerical equations of state for the $n=9$ SSMF model that the zero-temperature limit of the Kauzmann curve intersected the terminus of the liquid spinodal.  One must remember, however, that the focus of Ref. \cite{1644} was the Kauzmann locus whereas we strictly consider ideal glasses; the difference in the approaches is of course whether the total or configurational entropies are of concern.  Based on the discrepancy between the two studies, apparently there are subtle but fundamental differences in the implications the definitions of these two phenomena have for the analysis of liquids.  The present work therefore emphasizes
that an ideal glass transition does not necessarily correspond to a Kauzmann point; that is, the sudden confinement of a liquid to a unique, lowest-lying amorphous basin in its potential energy landscape can be a distinct occurrence from $S_\text{liquid}=S_\text{crystal}$.

Turning now to the pressure along the ideal glass transition locus, insertion of Eqs.\ \ref{ssmfscale1}-\ref{ssmfscale4} and \ref{avibapprox} into Eq.\ \ref{glassP} yields
\begin{equation}
\pis(\rho) =  \frac{n}{3} \gmin \rho^{n/3+1} - a \rho^2 
\label{ssmfpis}
\end{equation}
for the inherent structure contribution, and
\begin{eqnarray}
\pvib(\bg,\rho) & = & \frac{n+2}{2} \frac{\rho}{\bg(\rho)}
\nonumber \\
& = & \frac{n+2}{2C} \rho^{n/3+1}
\label{ssmfpvib}
\end{eqnarray}
for the vibrational part.  Without the functional form for $f[u]$, the expression for the pressure above $\tg$ is unspecified in the model.  Though thorough discussion of such specific expressions is beyond the scope of this paper, we present for the purposes of demonstration the equation of state formulated according to a Gaussian landscape (see Eq.\ \ref{gaussian}).  To give the most basic expression, we make the assumption of constant $g$, i.e., that the vibrational free energy is independent of basin depth.  It then follows from Eq.\ \ref{noglassP} that
\begin{eqnarray}
P(\beta,\rho) & = &\frac{n}{3} \rho^{n/3+1}
\left[ \ginf - \frac{(\ginf-\gmin)^2}{2 \sinf} \beta \rho^{n/3} \right]
\nonumber \\
& & - a \rho^2 + \frac{n+2}{2} \frac{\rho}{\beta} .
\label{ssmfgausP}
\end{eqnarray}
From this equation, one can see more clearly the possibility of a $T$-independent inherent structure equation of state.  The inherent structure pressure, given by all but the last term in \ref{ssmfgausP}, has a weak temperature dependence if the quantity $\sinf T$ is reasonably large.  In this case, the term in brackets is essentially given by the constant value $\ginf$ and the inherent structure pressure approaches
\begin{equation}
\pis \approx \frac{n}{3} \ginf \rho^{n/3+1} - a \rho^2 .
\label{highTssmfgausPis}
\end{equation}
Though the SSMF system is missing any dependence of the total number of basins on density, which would almost certainly accompany explicit attractive interactions, one could imagine this to have a relatively small effect at commonly-studied temperatures.  Qualitatively, then, the inherent structure pressure would continue to appear temperature-independent at high $T$.

\begin{figure}
\includegraphics[width=3.375 in]{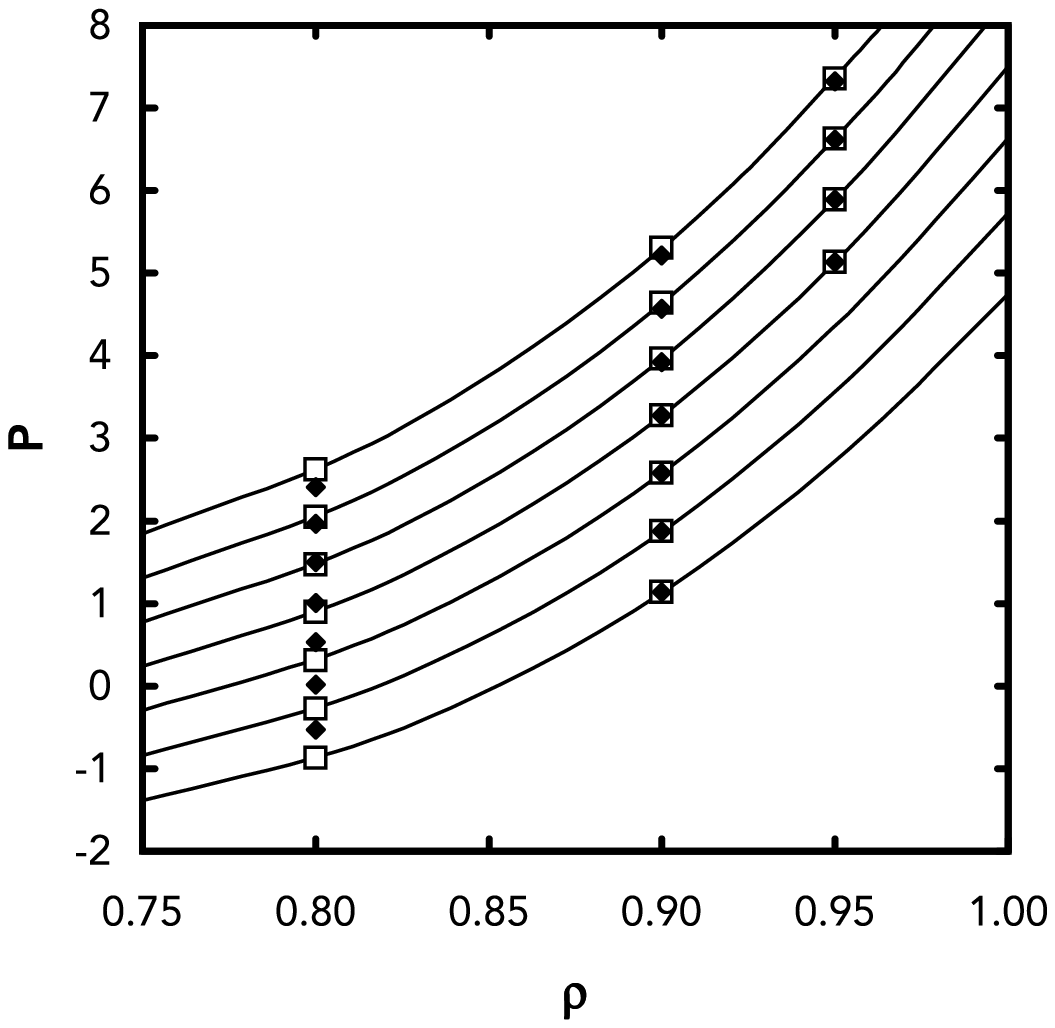}
\mycaption{Pressure isotherms for the landscape-derived equation of state of the soft-sphere/mean-field system, fitted to Lennard-Jones state points.  The predictions of the theory (lines) correspond to $T=0.7-1.3$ in increments of $0.1$, from bottom to top.  The filled diamonds are the simulation results from \cite{1783} and the squares are the corresponding predictions of the theory.  All variables are expressed in reduced Lennard-Jones units.}
\label{ljtheory}
\end{figure}

We have fitted the equation of state in \ref{ssmfgausP} to the
simulation results for the Lennard-Jones system in Ref.
\cite{1783}.  In Eq.\ \ref{ssmfgausP}, the parameters $\gmin$ and $\sinf$ are not
independent because they appear in the same coefficient.  In fitting the model,
we set $n=12$ and minimize the error in predictions for the pressure 
at several of the high-density, low-temperature state points in \cite{1783}, 
obtaining values of $4.64$, $16.5$, and $0.763$ for 
$\ginf$, $a$, and $(\ginf-\gmin)^2/\sinf$, respectively.
Our results are displayed in Fig. \ref{ljtheory}.  In general, Eq.
\ref{ssmfgausP} predicts the pressures well at increased densities,
which is the expected result since at these state points, the
Lennard-Jones system samples primarily the repulsive part of its
potential.  Interestingly, though, the predictions fare relatively
well for the two highest densities even at temperatures near the
critical temperature.  This is somewhat surprising considering 
the approximations in the theory, including the harmonic
approximation and the assumption of a basin-independent vibrational
free energy.  Using Eq.\ \ref{ssmfsastry}, we also obtain a Sastry
density of $0.76$ for this system; though this value differs from the
simulation result of $0.89$ in Ref. \cite{547}, one would expect a deviation
since the SSMF model neglects any explicit attractive interactions which are
sure to be important at $\rs$.

\section{Conclusions}
\label{conclusions}

We have presented a theoretical framework for liquids which incorporates the statistical properties of an energy landscape, is capable of describing an ideal glass transition, and corresponds with simulation studies performed in the inherent structure formalism.  In particular, the theory provides an explicit expression for the equation of state of a liquid and formally separates the pressure into vibrational and inherent structure components, both above and below the ideal glass transition.  Using the pressure separation, we have shown the presence of the Sastry density and its connection to the liquid spinodal.  Finally, we have used the theory to develop an elementary model of an energy landscape based on soft-sphere particles interacting with an additional mean-field attraction.  The model appears reasonably descriptive of simple liquids, suggesting the usefulness of a landscape-based reformulation of supercooled liquid thermodynamics.  An important conclusion from this work is that the ideal glass and Kauzmann loci are quite distinct for the soft-sphere model with mean-field attraction.  

Though the current work provides a detailed theoretical picture for energy landscapes, a number of questions remain.  Improved knowledge of the vibrational free energy at low temperatures is needed in order to provide a microscopic understanding of the distinction between an ideal glass transition and a Kauzmann equal-entropy point.  The apparent coincidence of the liquid spinodal with the Kauzmann locus remains unresolved; the results of our theory suggest that notions of an ideal glass at zero configurational entropy fail to capture what is occurring at this presumed convergence.  Finally, because our theory is based on the energy landscape, it contains information useful for both thermodynamics and kinetic phenomena (e.g., diffusion rates) \cite{932,1643}.  A useful future investigation might examine its ability to capture the slowdown of kinetic processes near the glass transition.

\begin{acknowledgments}
MSS gratefully acknowledges the support of the Fannie and John Hertz Foundation.  PGD gratefully acknowledges the support of the Department of Energy, Division of Chemical Sciences, Geosciences, and Biosciences, Office of Basic Energy Science (grant DE-FG02-87ER13714.)  FS and ELN gratefully acknowledge the support of MIUR FIRB and COFIN as well as INFM-PRA-GenFdT.
\end{acknowledgments}

\appendix*
\section{Identification of the inherent structure pressure}
\label{pressureappendix}

The particular separation of the pressure components in Eqs.\ \ref{pvibdef} and \ref{pisdef} is constructed to coincide with simulation studies of energy landscapes, in which inherent structures are found by energy minimization to local potential energy minima and their pressure determined from the usual virial expression.  The algorithmic procedure in these simulations may be thought of as an instantaneous quench to a non-equilibrium state at zero temperature.  Using a recently proposed thermodynamic formalism for such systems \cite{935}, we define the inherent structure pressure as the negative volume derivative of the free energy of the quenched state.  The non-equilibrium, per-particle free energy is given by
\begin{equation}
\ane(\bint,\bquench,\rho) = \phistar - \bint^{-1} \sigma(\phistar,\rho) + \avib(\bquench,\phistar,\rho)
\label{nefreeenergy}
\end{equation}
where $\bint$ refers to an internal temperature characterizing the out-of-equilibrium dynamics between inherent structures, $\bquench$ refers to the temperature to which the system is quenched, and $\phistar$ is the equilibrium value of $\phi$ prior to the quench, a function of the original temperature of the system $\beta$.  In the case where the quench temperature is absolute zero, the expression for $\bint$ is
\begin{eqnarray}
\bint & = & \pderiv{\sigma(\phistar,\rho)}{\phi} 
\nonumber \\
& = & \frac{\beta}{B(\beta,\phistar,\rho)} .
\label{bintdef}
\end{eqnarray}
Using this expression in the volume derivative of Eq.\ \ref{nefreeenergy} and taking the limit $T \to 0$ leads directly to the expression for inherent structure pressure in Eq.\ \ref{pisdef}.  The vibrational component of the pressure then follows from the remainder, $P-\pis$.  

An alternative but equivalent definition of the inherent structure pressure is the negative volume derivative of inherent structure energy at fixed configurational entropy, given by Eq.\ \ref{pisdef2}.  In the case of soft-spheres interacting through an inverse-power potential, one can demonstrate that Eq.\ \ref{pisdef2} is the correct $\pis$, that is, the one actually measured in simulations.  The unique property of soft-spheres is that upon isotropic compression in which all particle coordinates are simply rescaled, a configuration that is a potential energy minimum remains a minimum.  As a result, the number of members in a given ensemble of such configurations does not change with this type of volume deformation, and hence the same is true for the configurational entropy corresponding to any inherent structure.  Since the virial measures the response of the potential energy to an isotropic volume change of this sort and because of the constancy of configurational entropy during the compression, simulation pressure calculations for soft-sphere inherent structures are indeed given by Eq.\ \ref{pisdef2} and hence also \ref{pisdef}.

\bibliography{basin-refs-v6}

\figuresatend

\end{document}